\documentclass[a4paper,11pt]{article}
\usepackage[dvips]{graphicx}
\usepackage{amssymb}
\usepackage{amsmath}

\usepackage{cite}

\begin{document}

\title{A Criterion for Physically Acceptable Extra Dimensions with Boundaries}
\author{V. K. Oikonomou\thanks{
voiko@physics.auth.gr}\\
Technological Education Institute of Serres, \\
Department of Informatics and Communications 62124 Serres, Greece\\
and\\
Department of Theoretical Physics, Aristotle University of Thessaloniki,\\
Thessaloniki 541 24 Greece} \maketitle

\begin{abstract}
We present a criterion for deciding which compact extra
dimensional spaces yield physically reliable Newton's law
corrections. We study compact manifolds with boundary and without
boundary. The boundary conditions which we use on the boundaries
are Dirichlet or Neumann. We find that compact connected
Riemannian manifolds with Dirichlet boundaries are completely
excluded as extra dimensional spaces.
\end{abstract}

\section*{Introduction}

One of the main themes in the last century history of physics has
been the unification of all forces \cite{standard,standard1,standard2}. This direction in research was
motivated by the electromagnetism paradigm. Likewise, Kaluza and
Klein, independently worked towards the unification idea, and
tried to unify gravity and electromagnetism within general
relativity by adding a spatial extra dimension to our four
dimensional world. Nowadays, theories with extra dimensions have
refined representatives, such as superstring theories
\cite{clifford,pol,witten,mtheory1,b1,b1a,b1b,b2,scherk,c1,c2,c3,c3a,c4,c4a,leontarisbranes,leontarisbranes1}
and several UV extensions (see for example,
\cite{vafa,vafa1,donagi,donagi1,leontarisftheory,nanopoulos,nanopoulos1,foik1,foik2}).
Yet, string theories give predictions that energetically are way
above from the current experimental abilities that colliders
provide. On that account, the next step was string inspired low
energy effective quantum field theory which can be built on the
basic assumption that our world has extra spatial dimensions.
These extra dimensions should be compact and too small to be
(until now, 2011) detected \cite{arkani}. Actually, if the
compactification scale (that is, the inverse of the radius of the
compact dimension) is of the order TeV, then it is possible that
these extra dimensions could be experimentally detected in the
near future since their imprint may be present in many physical
laws and effects. In colliders, if a particle gains energy of the
order of the compactification scale, then it could be possible to
move along that extra dimensions, and this would imply lost energy
in our world. Along with collider experiments we expect to see the
imprint of extra dimensions indirectly and to other physical laws,
such as in Newton's law experiments.

\noindent The first correct scientific and mathematical
formulation of gravity is the Newton's law. It is very well
established and is, up to now, experimentally tested at distances
of the order $0.05{\,}mm$ \cite{Hoyle2006}. There exist
experiments that test Newtonian gravity such as the
E\"{o}tvos-type and also Cavendish type experiments. If extra
dimensions actually exist, it is possible that their imprint could
be found in the Newton's $1/r$ law when tested at small distances.
Many theoretical studies have been done towards this direction,
see for example
\cite{kehagias,kehagias1,leontaris1,leontaris2,randal,oikonomou3,oikonomou31,oikonomoukleidis}
and references therein. The form of the modifications of the
Newtonian potential, that the large compact extra dimensions
models predict is:
\begin{equation}\label{newton}
V(r)=-G_N\frac{MM'}{r}\Big{(}1+ae^{-r/\lambda}\Big{)},
\end{equation}
which are Yukawa type. In the above, $G_N$, $a$ and $\lambda$ are
the gravitational constant, strength and the range of the
gravitational force. The current experimental work constrains
these two parameters, concerning Yukawa type corrections.

\noindent Depending on the geometrical and topological features of
the extra dimensions, the strength and range will vary. The Yukawa
type corrections are essential to most large compact extra
dimensional spaces (but this does not apply to all compact spaces,
as we shall see) \footnote{Nonetheless, warped extra dimensional
models of Randall-Sundrum type \cite{randal}, give rise to
modifications of the form,
\begin{equation}\label{rs1}
V(r)=-G_N\frac{MM'}{r}\Big{(}1+\frac{C}{(kr)^2}\Big{)},
\end{equation}
with $C$ a constant inherent to the model (of order one) and $k$
the constant appearing in the warp factor of the metric (see also
\cite{reviewrs,reviewrs1} and for an alternative supersymmetric version see
\cite{friend})}.

\noindent In view of the E\"{o}tvos-type and also Cavendish type
experiments, it is necessary to know the form of corrections that
each compact space induces. There is a plethora of compact spaces
that can be extra dimensional spaces, which can vary in their
global topology, shape and their size.

\noindent However not all compact spaces give Yukawa type
corrections, for example the Dirichlet disk yields completely
unphysical Newton's law corrections (see \cite{oikonomoukleidis}).
These spaces have been used in the literature to solve several
phenomenological problems, however although interesting
phenomenologically, these must be excluded as extra dimensions
candidates.

\noindent It is necessary to have a criterion in order to
distinguish among all compact spaces, these that give physical
acceptable results. In this article we shall give such a criterion
for compact spaces with or without boundary. The later is less
involved in comparison to the former as we shall see, but we shall
present it for completeness.

\noindent This paper is organized as follows: We review the
technique for obtaining the Yukawa corrections to the Newton's law
in arbitrary dimensions. Next, we present the criterion that
determines which spaces can give reliable corrections to the
Newton's law. The conclusions follow at the end of the article.

\section*{Connection of Yukawa-type corrections with the zero modes of the Laplace-Beltrami operator}

We start with the very well known techniques for calculating
gravitational corrections to Newton's law (which is applicable to
all compact manifolds)
\cite{kehagias,oikonomou3,oikonomou31,oikonomoukleidis}. Based on reference
\cite{kehagias} we shall briefly present it.

\noindent Let $M^4\times M^n$ be the total spacetime, with $M^n$ a
n-dimensional compact manifold describing the extra space dimensions
and $M^4$ the four dimensional Minkowski spacetime. There exists a
complete set of orthogonal harmonic functions on $M^n$, $\Psi_m$,
satisfying the orthogonality condition:
\begin{equation}\label{orthogo}
\int_{M^n}\Psi_n(x)\Psi_m^*(x)=\delta_{n,m},
\end{equation}
and the completeness relation:
\begin{equation}\label{complet}
\sum_m\Psi_m(x)\Psi_m^*(x')=\delta^{(n)}(x-x').
\end{equation}
In the above the variable $x$ describes the compact space for now.
The harmonic functions $\Psi_m$ are eigenfunctions of the
$n$-dimensional Laplace-Beltrami operator $\Delta_n$ corresponding
to the manifold $M^n$, with eigenvalues $\mu_m^2$:
\begin{equation}\label{eigenvalueeq}
-\Delta_n\Psi_m=\mu_m^2\Psi_m.
\end{equation}
Let $V_{n+4}$, be the gravitational potential of the $n+4$
spacetime. It satisfies the Poisson equation in $n+3$ spatial
dimensions, when the Newtonian limit is taken:
\begin{equation}\label{lap}
\Delta_{n+3}V_{n+4}=(n+1)\Omega_{n+2}G_{n+4}M\delta^{(n+3)}(x)
\end{equation}
with $M$, the mass of the system, $G_{n+4}$ the Newton's
gravitational constant in $n+4$ dimensions and,
\begin{equation}\label{omega}
\Omega_{n+2}=\frac{2\pi^{\frac{n+3}{2}}}{\Gamma(\frac{n+3}{2})}.
\end{equation}
The large compact radius solution (\ref{lap}) is equal to:
\begin{equation}\label{neweq}
V_{n+4}=-\frac{G_{n+4}M}{r_n^{n+1}}.
\end{equation}
with $r_{n}$ the radius of the $n+3$-dimensional space in
spherical coordinates. Denoting by $r$ the radius of the
3-dimensional space of our world, then
$r_n=\sqrt{r^2+x_1^2+x_2^2+...+x_n^2}$. In our case the compact
space's radius is too small. Therefore we can write the harmonic
expansion of $V_{n+4}$ in terms of the eigenfunctions of the
product space $M^4\times M^n$, as follows:
\begin{equation}\label{harmonicexpansion}
V_{n+4}=\sum_m\Phi_m(r)\Psi_m(x),
\end{equation}
where $r$ denotes the coordinates describing $M^4$ and $x$ stands
for all the coordinates parameterizing $M^n$. Accordingly, the
$\Phi_m$ obey:
\begin{equation}\label{poiss3}
\Delta_3\Phi_m-\mu_m^2\Phi_m=(n+1)\Omega_{n+2}\Psi_m^*(0)G_{n+4}M\delta^{3}(x).
\end{equation}
The solution of equation (\ref{poiss3}) is equal to:
\begin{equation}\label{gravpot12}
\Phi_m(r)=-\frac{\Omega_nG_{n+4}M\Psi_m^*(0)}{2}\frac{e^{-|\mu_m|r}}{r}.
\end{equation}
and lastly the gravitational potential reads:
\begin{equation}\label{fingravpot}
V_{n+4}=-\frac{\Omega_nG_{n+4}M}{2r}\sum_m\Psi_m^*(0)\Psi_m(x)e^{-|\mu_m|r}.
\end{equation}
The physical quantities of 4-dimensional point particles have no
coordinate dependence on the internal compact space $M^n$, thus we
take $x=0$ in (\ref{fingravpot}). This way the four dimensional
gravitational potential can be cast as:
\begin{equation}\label{fingravpot1}
V_{4}=-\frac{G_{4}M}{r}\sum_m\Psi_m^*(0)\Psi_m(0)e^{-|\mu_m|r},
\end{equation}

\section*{Not All Compact Spaces Yield Physically Acceptable Results}

\noindent As it is obvious from equation (\ref{fingravpot1}), the
Yukawa type corrections are obtained when the compact space has a
zero mode. On that account, the most dominant contribution to the
gravitational potential comes from the first Kaluza-Klein (KK)
excitation of the extra compact space (or equivalently from the
lowest non-zero eigenvalue of the Laplace-Beltrami operator). Most
compact spaces do have a zero mode and therefore give the expected
Yukawa type correction to the potential.

There are however examples of extra dimensional spaces that yield
completely unphysical results, as we already mentioned previously.
For example, as is found in reference \cite{oikonomoukleidis},
when the extra space is a Dirichlet disk (that is, a disk with the
extra dimensional field satisfying Dirichlet boundary conditions
on the boundary), the correction to the potential is:
\begin{equation}\label{tetanusfree}
V_{4}\simeq -\frac{G_{4}M}{r}\sum_ne^{-\frac{x_{0n}}{R}r}.
\end{equation}
with $x_{0n}$ the $n$-th root of the $J_0(x)$ Bessel function. By
keeping the lowest eigenvalue in the above sum (which is
$x_{01}=2.40483$) we have:
\begin{equation}\label{butpainremains}
V_{4}\simeq -\frac{G_{4}M}{r}e^{-\frac{2.40483}{R}r}.
\end{equation}
The above form is a Yukawa gravitational potential and not a
Yukawa correction to the gravitational potential. As noted in
\cite{oikonomoukleidis} this form of potentials can also be found
in the case the extra space is a hyperbolic manifold (a similarity
that is probably accidental).

\noindent Although not frequently used in the literature, extra
dimensions with Dirichlet or Neumann boundaries provide very
interesting phenomenological features to field theory models with
large extra dimensions. In particular the Dirichlet disk was used
in reference \cite{papineau} in a phenomenological six dimensional
model, in which strong coupling is related with massless or light
degrees of freedom in the spectrum. This is a frequent phenomenon
in string theory and the authors of \cite{papineau} achieved to
construct a field theory model in which massless or very light
degrees of freedom emerge upon turning on small couplings in the
theory (the last initially had only heavy degrees of freedom in
the spectrum). In the same way, the authors of \cite{papineau1}
used the Neumann disk, in addition to the Dirichlet one.

\noindent Obviously, the spaces with Dirichlet boundaries yield
completely unphysical Newton's law corrections. Therefore, even if
the phenomenology these compactifications provide is quite
interesting, these must be excluded from extra dimensions
searches. Contrarily, if we use Neumann boundary conditions on the
boundary circle, the result is different, since the corrections to
the gravitational potential take the Yukawa type form
\cite{oikonomoukleidis}.

\noindent Note that the absence of a zero mode in the Dirichlet
case and the presence of a zero mode in the Neumann case, is
responsible for these results. We shall see that this is by far
not accidental. Specifically, the Dirichlet problem and the
Neumann problem are actually interrelated. Therefore, we
generalize the criterion for manifolds without boundary to the
case the manifolds have Dirichlet or Neumann boundaries. We focus
on the case that the boundary conditions are Neumann or Dirichlet.
The compact space may have arbitrary topology but is assumed to be
Riemannian.

\section*{A Quick Mathematical Review}

In order to make the article self-contained, we briefly quote here
some of the definitions of the well known mathematics we shall
use. We start with the inclusion map, which is defined between a
space $M$ and an open subset $A$ of it, as follows:
\begin{equation}\label{inlusionm}
i:A\rightarrow M,
{\,}{\,}{\,}{\,}{\,}{\,}i(a)=a{\,}{\,}\mathrm{for}{\,}{\,}a{\,}
\epsilon{\,}{\,}{\,}A
\end{equation}
Let $M$ be a $n$-dimensional smooth Riemannian manifold. The
exterior derivative $d$ maps a form $\omega$ that belongs to the
space of smooth $p$-forms $\Omega (M)^p$ to the space of
$p+1$-forms, that is:
\begin{equation}\label{exterionr}
d:\Omega(M)^{p}\rightarrow \Omega(M)^{p+1}.
\end{equation}
The Hodge operation $\star$ is defined as the linear map,
\begin{equation}\label{hodgestar}
\star :\Omega(M)^{p}\rightarrow \Omega(M)^{m-p}.
\end{equation}
The adjoint of $d$, denoted as $\delta$, maps $p$-forms to
$p-1$-forms, that is:
\begin{equation}\label{adjointexterior}
\delta :\Omega(M)^{p}\rightarrow \Omega(M)^{p-1}.
\end{equation}
We define the operator $\delta$:
\begin{equation}\label{defeadj}
\delta =(-1)^{mr+m+1}\star d \star.
\end{equation}
Using the operators $d$ and $\delta$, we define the Laplacian
$\Delta_p =d\cdot \delta +\delta \cdot d$, which is a map:
\begin{equation}\label{laplacian}
\Delta_p :\Omega(M)^{p}\rightarrow \Omega(M)^{p}.
\end{equation}

\section*{Zero Modes of the Laplace-Beltrami Operator, Betti Numbers and the Poincare-Lefschetz Duality}

\noindent As we mentioned earlier, the zero modes of the
Laplace-Beltrami operator corresponding to the extra dimensional
space and the existence of Yukawa-type corrections due to compact
space are strongly related. Indeed if no zero modes exist, then
the compact space does not give the expected gravitational
potential form.

\subsubsection*{The trivial case-Manifolds without boundary}

\noindent Let $M^n$ be a compact $n$-dimensional space which is
the extra dimensional space to $M^4$ (we follow the notation of
the previous sections). We suppose that $M^n$ is compact, oriented
and Riemannian. Then we can define the de-Rham complex and
therefore cohomology classes on this manifold. Hence we expect all
the known cohomology theorems to hold. Let $\Omega^p(M^n)$ be
space of p-forms, $\omega$. The cohomology class $H^p(M^n)$ counts
the forms that are closed closed but not exact, under the exterior
derivative $d$ and it's dual $\delta$. One can define the
Laplacian $\Delta_p=d\delta+\delta d$ acting on $p$-forms (the
Laplacian coincides with the Laplace-Beltrami operator when
applied to 0-forms). It is very well known that on every compact
Riemannian manifold without boundary, the following holds:
\begin{equation}\label{coho1}
\mathrm{dimker}(\Delta_p)=\mathrm{dim}H^p(M^n).
\end{equation}
Thereupon, the p-th Betti number, $b_p$, is equal to:
\begin{equation}\label{betti1}
b_p=\mathrm{dim}H^p(M^n).
\end{equation}
Thus, it is obvious that $b_p=\mathrm{dimker}\Delta_p$. Therefore,
finding zero modes of the Laplace-Beltrami operators is equivalent
to finding zero modes of 0-forms. Consequently we can trivially
conclude that zero modes of the Laplace-Beltrami operator exist
whenever the $b_0$ Betti number is non-zero. Particularly, the
$b_0$ Betti number counts the number of different disconnected
component of the manifold $M^n$. Hence if the manifold is
connected (which is the usual case in studies of large extra
dimensions), $b_0=1$. Indeed all the known compact Riemannian
manifolds without boundary, have $b_0=1$, such as the n-torus, the
sphere and so forth.

\subsubsection*{Manifolds with Boundary and Relative Cohomology}

\noindent Consider now that the compact manifold $M^n$ is as
before an $n$-dimensional oriented and Riemannian but with a
smooth boundary. We want again to solve the problem of finding
when there exist zero modes of the Laplace-Beltrami operator, with
the 0-form $\omega$ obeying Dirichlet or Neumann boundary
conditions on the boundary. Thus the problems at hand are:
\begin{align}\label{diriclap}
&\Delta \omega=0
\\ \notag & \omega |_{\partial M^n}=0
\end{align}
for the Dirichlet case, and,
\begin{align}\label{neulap}
&\Delta \omega=0
\\ \notag & \frac{\partial \omega}{\partial x} \Big{|}_{\partial M^n}=0
\end{align}
for the Neumann case. Again, $x$ denotes the coordinate describing
the internal space $M^n$. Let us generalize to the case that
$\omega $ is a $p$-form. As before, the space of smooth $p$-forms
is $\Omega^p (M^n)$. The harmonic $p$-forms $\omega$ are elements
of $\mathrm{Harm}^p(M^n)$ and satisfy, $d\omega =0$ and $\delta
\omega =0$ (or simply $\Delta \omega =0$). Let $i:\partial
M^n\rightarrow M^n$ be the embedding (the inclusion map). Then the
$p$-forms that are harmonic forms and satisfy Dirichlet boundary
conditions, belong to the space $\mathrm{Harm}_D^p(M^n,\partial
M^n)$ (the last is known as relative cohomology) which consists of
forms that formally satisfy:
\begin{align}\label{diriclapfor}
&\Delta \omega=0
\\ \notag & i^*\omega =0.
\end{align}
The map $i^*$ is the pullback of the embedding $i$. It is not
difficult to see that the problem (\ref{diriclapfor}) reduces to the
one described by (\ref{diriclap}) when 0-forms are considered. In
the Neumann case, the p-form Neumann problem reads,
\begin{align}\label{neulap}
&\Delta \omega=0
\\ \notag & i^*(\star \omega) =0.
\end{align}
and the space of $\omega's$ that satisfy the above, is denoted by
$\mathrm{Harm}_N^p(M^n,\partial M^n)$. In the above ''$\star$'' is
the well known Hodge operator. The two spaces
$\mathrm{Harm}_N^p(M^n,\partial M^n)$ and
$\mathrm{Harm}_D^p(M^n,\partial M^n)$ are interrelated by the
Poincare-Lefschetz duality \cite{theorem} that states:
\begin{equation}\label{poincarelef}
\mathrm{dimHarm}_N^p(M^n,\partial
M^n)=\mathrm{dimHarm}_D^{n-p}(M^n,\partial M^n)=b_p(M^n),
\end{equation}
with $n$ the dimension of the space $M^n$ and $b_p$, the $p$-th
Betti number. Let us see the impact of this duality on the zero
modes of 0-forms that satisfy Dirichlet boundary conditions. We
must find the quantity $\mathrm{dimHarm}_D^{0}(M^n,\partial M^n)$,
thus we set $p=n$ in (\ref{poincarelef}) and we have:
\begin{equation}\label{poin}
\mathrm{dimHarm}_N^n(M^n,\partial
M^n)=\mathrm{dimHarm}_D^{0}(M^n,\partial M^n)=b_n(M^n).
\end{equation}
In the case we are interested in Neumann 0-forms, we must set
$p=0$ in (\ref{poincarelef}) and thus,
\begin{equation}\label{poin1}
\mathrm{dimHarm}_N^0(M^n,\partial
M^n)=\mathrm{dimHarm}_D^{n}(M^n,\partial M^n)=b_0(M^n).
\end{equation}
Relations (\ref{poin}) and (\ref{poin1}) are really useful.
Firstly we can see that the problem of finding the number of zero
modes of the Laplace-Beltrami operator for compact spaces with
boundaries simply reduces to a problem of algebraic topology, with
the last being finding the Betti numbers of the manifold $M^n$.
Secondly the Dirichlet and Neumann problems are related. In
particular, the number of zero modes that the Dirichlet
Laplace-Beltrami operator has, is equal to the zero modes that the
$n$-forms of the Neumann Laplace-Beltrami operator have and
conversely.

\noindent Having the above results at hand we apply them in the
case of the Neumann and Dirichlet disk, that we discussed in the
previous section. For the disk manifold we have that $b_0(M^2)=1$
and $b_2(M^2)=0$, thus $\mathrm{dimHarm}_N^0(M^2,\partial M^2)=1$
and $\mathrm{dimHarm}_D^0(M^2,\partial M^2)=0$. It is obvious that
no zero modes exist in the Dirichlet case while in the Neumann,
exists one. In conclusion we sum up:

\begin{itemize}

\item Since any contractible space has trivial homology (which
implies that $b_0=1$ and all the other Betti numbers are equal to
zero), the homological trivial (with boundary) spaces are not
likely to be extra dimensional spaces when Dirichlet boundary
conditions are used on their boundary.

\item Compact spaces for which the Laplacian of 0-forms on these
has no zero modes, are not likely to be extra dimensional spaces.

\end{itemize}

\section*{Conclusions}

Most known compact extra dimensional spaces yield Yukawa-type
corrections to the $1/r$ four dimensional gravitational law.
However not all compact spaces give reliable results, for example
compact spaces with Dirichlet boundaries. In this article, we
addressed the problem of finding a rule that tells us, which
spaces can give reliable corrections to the 4-dimensional
gravitational potential. We found that the problem is reduced to
an algebraic topology one, since the number of the
Laplace-Beltrami operator's zero modes is related to the Betti
numbers of the compact space. For manifolds without boundaries,
all the connected compact spaces have a zero mode when the first
Betti number is equal to, $b_0=1$. Recall that the first Betti
number counts the disconnected components of the extra dimensional
manifold. Still, extra dimensional spaces must be connected for
consistency. So all compact connected boundary-less manifolds can
yield Yukawa-type corrections.

\noindent In the case that manifolds with boundaries are
considered, things are somehow different. We found that connected
compact manifolds with Dirichlet boundaries must be excluded from
studies of extra dimensional physics. We related this result to
the Betti number of the manifold. On the contrary, using similar
arguments, we found that the spaces with Neumann boundaries always
give the expected form of corrections to the gravitational
potential, that is Yukawa-type form.

\noindent The exclusion of some spaces as extra dimensional
spaces, can make the identification of extra dimensions somehow
easier. However in theories with extra dimensions there is a
problem that can make the identification of the exact topology and
geometry of the extra dimensions very difficult. It is known as
the shadowing effect (see for example
\cite{dienes,oikonomoushape}). In some cases even the number of
extra dimensions will be difficult to determine because results
may coincide. In view of such difficulties in order to have a
clear picture of the experimental results, we must combine the
results coming from other physics areas, such as cosmic ray
developments, \cite{Nikolaidis,erdas,Nikolaidis1}, Casimir effect
constraints
\cite{odintsov,odintsov1,odintsov2,elizalde,elizalde1,elizalde2,elizalde3,elizalde4,elizalde5,elizalde6,oikonomou1,oikonomou11,oikonomou12,oikonomou13,kirsten,mostepanenko,perivolaropoulos}
and alternative scenarios
\cite{cullen,kehagias,leontaris2,direct,direct1,direct2,vergados,ued,oikonomou2,oikonomou21,oikonomou22,caposielo,kaloper}.
These scenarios, take into account compact extra dimensions and
can pose restrictions on the number and even the size of the extra
dimensions.

\end{document}